\documentclass{elsart}
\usepackage{graphicx}
\usepackage{tabularx}
\journal{Surface Science}

\begin{document}
\begin{frontmatter}

\title{Formation of a quasicrystalline Pb monolayer on the ten-fold surface of the decagonal Al-Ni-Co quasicrystal}

\author[LivPh]{J.A.~Smerdon},
\author[LivPh]{L. Leung\thanksref{now}},
\thanks[now]{Present address: Department of Chemistry, University of Toronto, Toronto, Ontario M5S 3H6, Canada}
\author[LivPh]{J.K. Parle},
\author[Ames]{C.J. Jenks},
\author[LivPh]{R. McGrath\corauthref{cor}}
\corauth[cor]{Corresponding author.}
\ead{mcgrath@liv.ac.uk}
\author[Nancy]{V. Fourn\'{e}e},
\author[Nancy]{J. Ledieu}
\address[LivPh]{Department of Physics and Surface Science Research Centre, The University of Liverpool, Liverpool L69 3BX, UK}
\address[Ames]{Ames Laboratory, Iowa State University, Ames, IA 50011, USA}
\address[Nancy]{LSG2M, CNRS UMR 7584, Ecole des Mines, Parc de Saurupt, 54042 Nancy Cedex, France}

\begin{abstract}

Lead has been deposited on the ten-fold surface of decagonal Al$_{72}$Ni$_{11}$Co$_{17}$ to form an epitaxial quasicrystalline single-element monolayer. The overlayer grows through nucleation of nanometer-sized irregular islands and the coverage saturates at 1 ML. The overlayer is well-ordered quasiperiodically as evidenced by LEED and Fourier transforms of STM images. Annealing the film to 600 K improves the structural quality, but causes the evaporation of some material such that the film develops pores. Electronic structure measurements using X-ray photoemission spectroscopy indicate that the chemical interaction of the Pb atoms with the substrate is weak.
\end{abstract}

\begin{keyword}
STM, LEED, epitaxy, quasicrystal, lead
\PACS 61.44.Br, 68.35.Bs, 68.37.Ef
\end{keyword}
\end{frontmatter}
\newpage

\section{Introduction}

Quasicrystals are metallic alloys that exhibit exotic kinds of structuring and symmetries that are not found in periodic crystals.  The most common kinds of quasicrystal can exhibit five-fold (icosahedral) or ten-fold (decagonal) rotational symmetries as evidenced using diffraction techniques, indicating that these materials are well-ordered but aperiodic materials.  A well-known aperiodic mathematical analogue to a quasicrystal is the Penrose tiling, which has previously been successfully mapped onto scanning tunnelling microscopy images from various quasicrystals \cite{Ledieu01-SSL,Sharma04}.

These unusual materials have been the subject of intense study for the past two decades in a concerted attempt to unravel the mysteries of their structure and properties. The technologically desirable combination of high hardness coupled with low surface energy (leading to low friction) and other unusual properties, for example a negative coefficient of electrical resistivity, has attracted much interest from research groups around the world \cite{Trebin03}.  The development in recent years of the ability to prepare flat quasicrystal surfaces suitable for surface science techniques has led to an upsurge in studies of clean quasicrystal surfaces and  their interactions with adsorbing atomic and molecular species.

The progress that has been achieved in understanding and classifying atomic adsorption on quasicrystal surfaces has been summarised in some recent review papers \cite{McGrath02-JPCM,Fournee04,Sharma07,Smerdon08-PM}. Several atomic species have been found to chemisorb and subsequently form epitaxial complete layers and multilayer films. These can be arranged into a number of broad classifications. One class contains elements which form crystalline structures which are aligned along substrate high symmetry directions, with or without some intermixing, e.g. Fe adsorbed on $i$-Al-Pd-Mn \cite{Wearing07} or Fe adsorbed on $d$-Al-Ni-Co \cite{Wearing08}. Another class of adsorbates can be described as forming modulated multilayers. Such structures have been found for Cu adsorbed on the five-fold surface of icosahedral Al-Pd-Mn ($i$-Al-Pd-Mn) \cite{Ledieu04} and for Co adsorbed on both $i$-Al-Pd-Mn and the ten-fold surface of decagonal Al-Ni-Co ($d$-Al-Ni-Co) \cite{Smerdon06}. They consist of small domains of crystalline material oriented in five directions corresponding to high symmetry orientations on the substrate. The domains themselves have a quasiperiodic structural modulation consistent with terms of the Fibonacci sequence \cite{Ledieu04,Ledieu05,Smerdon06a}.

A third class are systems where the adsorbing species adopts the structure of the quasicrystal substrate. Such systems have only been discovered to date with coverages from low sub-monolayer up to saturated monolayer. Adsorption of Sb and Bi on both $i$-Al-Pd-Mn and $d$-Al-Ni-Co \cite{Franke02} has been found to lead to the formation of  monolayer pseudomorphic systems. Aluminium atoms form small pentagonal clusters (`starfish') on $i$-Al-Cu-Fe at low sub-monolayer coverages of $\approx0.04$ ML, although at higher coverages disordered adsorption occurs \cite{Cai03}. This has also been found for Si adsorption on both $i$-Al-Pd-Mn \cite{Ledieu06} and $d$-Al-Ni-Co \cite{Leung06}. Xenon adsorbed on the ten-fold surface of $d$-Al-Ni-Co was found to physisorb, with the strength of the adsorbate-adsorbate and adsorbate-substrate interactions being comparable. The first layer had five-fold symmetry, and a transition to six-fold symmetry indicating bulk Xe formation was observed after a coverage of one monolayer \cite{Curtarolo05,Diehl06,Diehl07,Setyawan06,Setyawan07}.

Recently Pb has been added to the list of adsorbed species which form pseudomorphic overlayers \cite{Ledieu08}. When dosed on the five-fold surface of $i$-Al-Pd-Mn at low sub-monolayer coverages, Pb atoms were found to nucleate into pentagonal structures. As the atomic density of the atoms increased, an interconnecting network was formed until a saturated quasicrystalline monolayer was observed. Using scanning tunneling spectroscopy and ultraviolet photoemission spectroscopy, this Pb monolayer was found to display an electronic pseudogap similar to those found in bulk quasicrystalline materials \cite{Stadnik01}.

In this paper, we describe the results of experiments probing the adsorption of Pb, this time on the ten-fold surface of $d$-Al-Ni-Co. This surface has been shown to be a relaxed truncated bulk structure,
having the same composition as the bulk. The outermost layer spacing is contracted by 10\% relative to the bulk interlayer spacing, while the next layer spacing is expanded by 5\%. A small degree of intralayer rumpling is observed within each layer \cite{Cox01,Ferralis04}. This surface is chemically and structurally distinct from that of $i$-Al-Pd-Mn, and therefore it is of considerable interest to investigate whether the quasiperiodic ordering found for $i$-Al-Pd-Mn/Pb is repeated in this system.

\section{Experimental details} \label{exper}

 The Al$_{72}$Ni$_{11}$Co$_{17}$ quasicrystal samples, produced at Ames laboratory using the melt decantation method, were polished successively with 6 $\mu$m, 1 $\mu$m and 1/4 $\mu$m diamond paste before introduction to vacuum and thereafter was prepared in cycles consisting of 45 minutes sputtering with 3 keV Ar$^{+}$ ions followed by 4 hours annealing to 1070 K, using electron-beam heating, up to a total annealing time of 20 hours.  Following this preparation, low energy electron diffraction (LEED) patterns had well defined peaks and impurities were undetectable by Auger electron spectroscopy.

The scanning tunnelling microscopy (STM) studies of adsorption and the Auger electron spectroscopy (AES) measurements were carried out in an Omicron variable temperature STM (VT-STM) chamber at the University of Liverpool.  The base pressure of the system was 1$\times$10$^{-10}$ mbar.  The Pb evaporation was performed using an Omicron EFM-3 electron-beam evaporator, with an ion flux monitor reading of 120 nA giving approximately 0.2 ML coverage after 15 minutes.  The chamber pressure did not exceed 2.5$\times$10$^{-10}$ mbar during evaporation.

The x-ray photoemission spectroscopy (XPS) measurements and LEED results were obtained at the Ecole des Mines, Nancy, France.  STM and AES measurements were also carried out for comparative purposes. In those experiments the Pb source was calibrated by means of XPS and STM using an Al(111) crystal. Annealing was done at a temperature of 653 K.

\section{Results and analysis}

\begin{figure}[!t]
 \begin{center}
\includegraphics[width=0.9\textwidth]{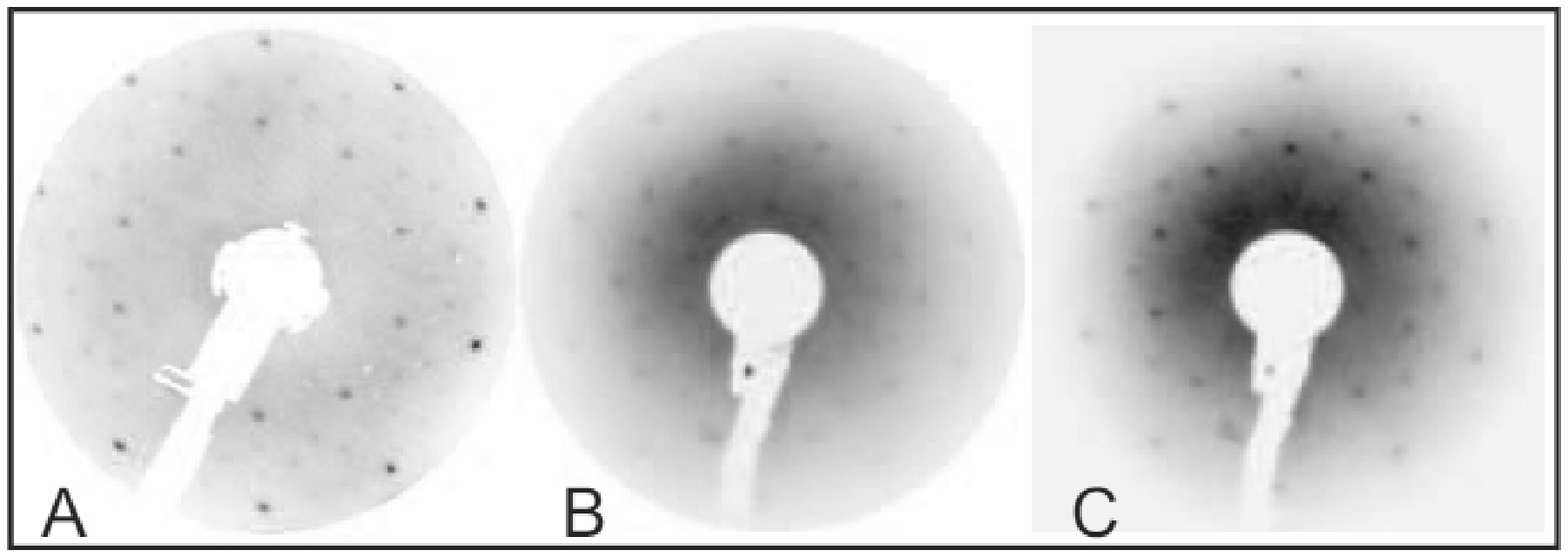}
 \caption{LEED patterns recorded at 75 eV on (a) the clean surface, (b) with 1ML Pb as deposited on the $d$-Al-Ni-Co quasicrystal surface and (c) after annealing this film to 653 K.}
  \label{LEED1}
 \end{center}
 \end{figure}

Fig. \ref{LEED1}(a) shows LEED data recorded at 75 eV from the clean surface after the preparation procedure described in Section \ref{exper}. Quasicrystals are aperiodic materials, which means that reciprocal space is infinitely dense; in practice LEED patterns are dominated by the most intense reflections. In this case several rings of ten spots are observable. The radii of the two most intense rings are related by a factor $\tau$ \footnote{The number $\tau=\frac{1+\sqrt{5}}{2}=1.618...$, known as the golden ratio, is intrinsic to the geometry of pentagons, to Penrose tilings and the Fibonacci sequence.}; this is indicative of quasiperiodic ordering. The  surface has a step-terrace morphology with terrace widths of order 20 nm as observed using STM. STM images yield a dense ten-fold fast Fourier transform (FFT) (not shown), again indicative of quasiperiodic ordering.

\begin{figure}[!t]
\begin{center}
\includegraphics[width=0.75\textwidth]{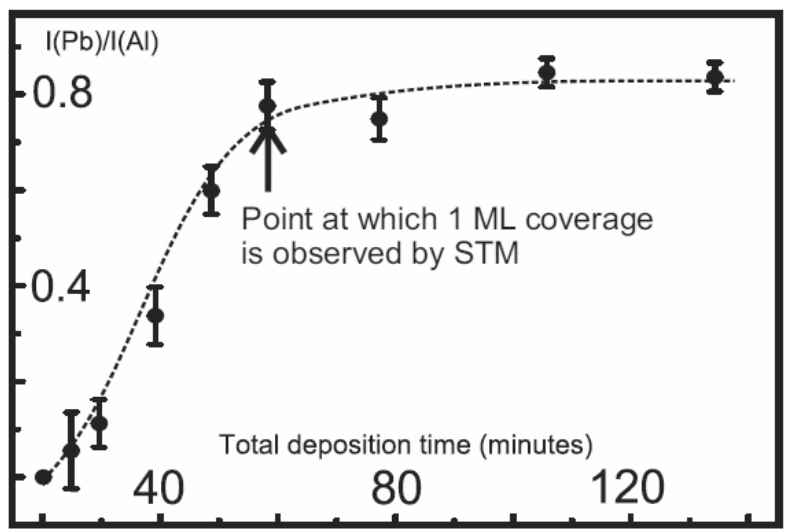}
\caption{Plot of Pb(NOO)/Al(LMM) Auger peak intensity ratio versus deposition time for 1 ML of Pb adsorbed on the ten-fold surface of $d$-Al-Ni-Co.}
\label{auger}
\end{center}
\end{figure}

Figure \ref{auger} shows AES measurements of the film during deposition. The ratio of intensities of Pb(NOO) and Al(LMM) peaks is plotted as a function of coverage. Following an initial rapid uptake, the Pb coverage saturates at monolayer coverage. This behaviour was also observed for $i$-Al-Pd-Mn/Pb, and is discussed further in Section \ref{discussion}.

\begin{figure}[!t]
\begin{center}
\includegraphics[width=0.75\textwidth]{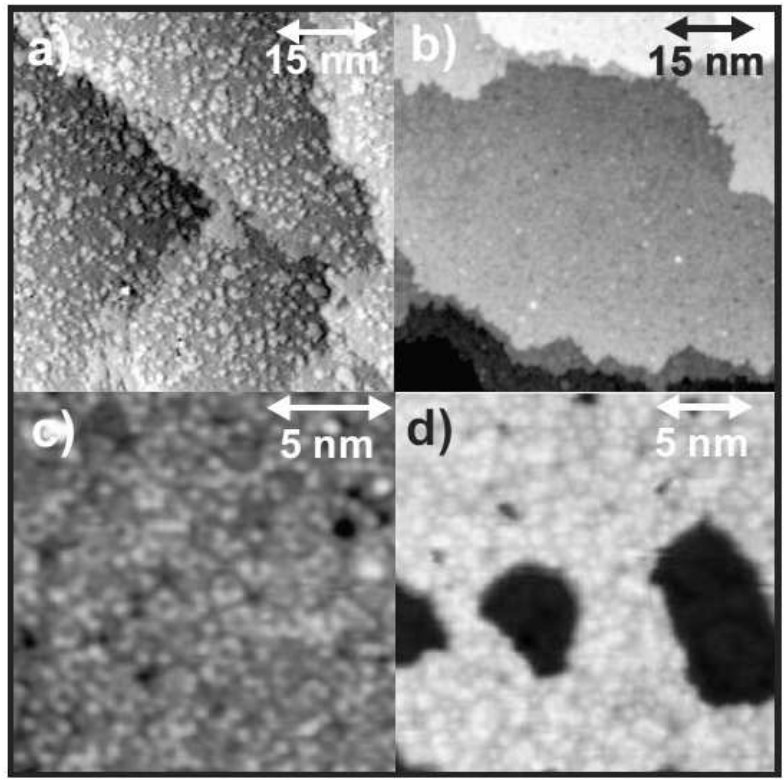}
\caption{Scanning tunnelling microscopy images of the formation of the Pb film on \emph{d}-Al-Ni-Co. (\emph{a}); 57 nm $\times$ 57 nm, 0.32 ML; (\emph{b}); 64 nm $\times$ 64 nm, 1 ML; (\emph{c}); 15 nm $\times$ 15 nm, 1 ML, showing structural motifs associated with quasicrystalline ordering; (\emph{d}); 20 nm $\times$ 20 nm following annealing to 600 K, showing the porosity developed in the film.}
\label{pbgrowth}
\end{center}
\end{figure}

An overview of the growth of the Pb film is shown in Fig. \ref{pbgrowth}.  Figure \ref{pbgrowth}(a) shows the system after deposition of 0.3 ML of Pb. The Pb atoms can be seen to nucleate in small clusters on the surface of average size $\approx$ 1 nm. During growth of the film, noise at the edges of Pb islands evidenced the diffusion of Pb adatoms on the surface.  Resolution within the islands themselves was poor.

After further evaporation to a coverage of 1 ML, the surface was completely covered with Pb atoms as shown in  Fig. \ref{pbgrowth}(b). At this coverage, atomic resolution was routinely obtained. A portion of a terrace of dimensions 15 nm $\times$ 15 nm is shown in Fig. \ref{pbgrowth}(c). The bright protrusions are interpreted as Pb atoms.  The LEED pattern after adsorption of 1 ML is shown in Figure \ref{LEED1}(b) for a beam energy of 75 eV. Again the  diffraction spots are quasiperiodically spaced with a characteristic $\tau$-scaling relationship between the rings indicative of quasiperiodic ordering. The pattern is  similar to that observed for the clean surface, although some of the fainter spots in Fig. \ref{LEED1}(a) are no longer visible.

Figure  \ref{pbgrowth}(d) shows the 1 ML coverage after annealing to a temperature of 600 K for 15 minutes. The film also develops nanosized `pores' as shown in  Fig. \ref{pbgrowth}(d). These pores could be due to evaporation of Pb atoms from this surface or diffusion of Pb into the substrate. There could even be diffusion of substrate atoms through the Pb film followed by evaporation although this is unlikely at the annealing temperature chosen. Further experiments are needed to clarify this point. The LEED pattern recorded from this phase is shown in Fig. \ref{LEED1}(c). Compared to the unannealed surface, the intensity of the spots has increased, suggesting an improvement in structural quality. There are also changes in the relative intensities of some of the rings, suggesting some structural changes may take place upon annealing.

\begin{figure}[!b]
\begin{center}
\includegraphics[width=0.9\textwidth]{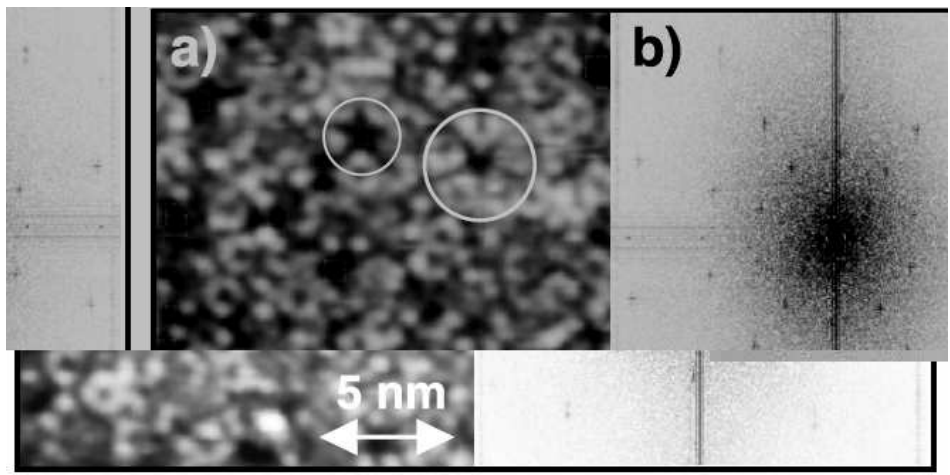}
\caption{(a) Quasicrystalline Pb film on $d$-Al-Ni-Co (15 nm $\times$ 15 nm). The circles indicate prominent pentagonal motifs. (b) FFT of a 50 nm $\times$ 50 nm image from this film, showing several decagonal rings indicating excellent quasiperiodic ordering.}
\label{tauscale}
\end{center}
\end{figure}

Fig. \ref{tauscale}(a) shows a high-resolution STM image of the Pb monolayer film. A five-fold pentagon (defining a five-fold hollow) is circled. Such hollows are a very distinctive feature of clean quasicrystalline surfaces; on the clean surface of $i$-Al-Pd-Mn they have been dubbed `dark stars' \cite{Papadopolos02,Krajci06c}. Also circled is another larger five-fold feature which is reminiscent of structures found on the five-fold surfaces of icosahedral quasicrystals \cite{Papadopolos02} and on the clean surface of $d$-Al-Ni-Co \cite{Ferralis04}. The observation of these features indicates that the structure of the quasicrystalline overlayer has similarities to that of the clean surface. The distance between the bright protrusions forming the pentagonal hollow is 4.9 \AA{}, similar to the size of the pentagons observed for Al-Pd-Mn/Pb. Fig. \ref{tauscale}(b) shows an FFT from the surface shown in (a), obtained from an image 50 nm $\times$ 50 nm in dimensions. Three distinct rings of spots are observed; again the radii are $\tau$-scaled.

\begin{figure}[!t]
\begin{center}
\includegraphics[width=0.7\textwidth]{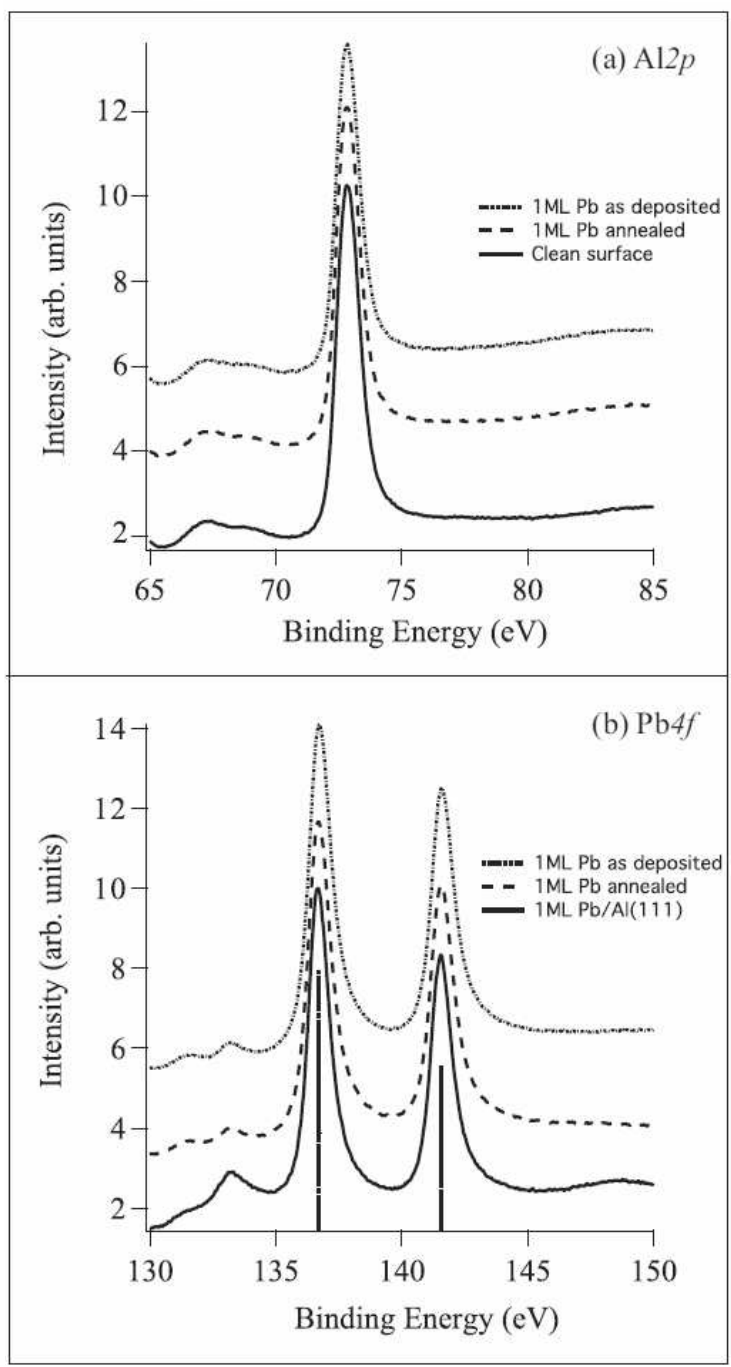}
\caption{(a) XPS spectra of the Al 2\textit{p} core levels for the clean \textit{d}-Al-Ni-Co surface (solid line), for 1 ML of Pb as deposited (dotted line), and 1 ML of Pb annealed to 653 K (dashed line); (b) XPS spectra of the Pb 4\textit{f} core levels for 1 ML of Pb adsorbed on Al(111) (solid line), 1 ML of Pb as deposited on the \textit{d}-Al-Ni-Co surface (dotted line) and 1 ML after annealing to 653 K (dashed line). The markers at 136.6 eV and 141.5 eV indicate the position of the core-levels for elemental Pb.}
\label{corelevel}
\end{center}
\end{figure}

The Al $2p$ and Pb $4f$ core levels were also recorded upon adsorption of a monolayer of Pb. Fig. \ref{corelevel}(a) shows that the shape of the Al $2p$ core level measured is identical to that of the clean quasicrystal surface for the monolayer as deposited at room temperature or annealed to 653 K. The Pb $4f$ core level recorded from the quasiperiodic Pb monolayer has an identical shape to that measured from one 1ML of Pb grown on Al(111) (Fig. \ref{corelevel}(b)). The lack of new components and chemical shifts within the core level peaks is consistent with the immiscibility of Pb and Al \cite{Johnson01,Matolin06}. No changes were detected in the Ni 2$p$ core level either. A comparison of the intensities of the Pb and Al photoemission peaks indicates that the density of Pb atoms is 0.09 atoms/\AA{}$^2$, identical to that found on the $i$-Al-Pd-Mn surface and on the Al(111) surface. For comparison, the value found experimentally using LEED for the clean surface is 0.123 atoms/\AA{}$^{2}$ \cite{Ferralis04}.

\section{Discussion} \label{discussion}

The most important result presented is the observation of quasiperiodic ordering of Pb on the ten-fold $d$-Al-Ni-Co surface. It was not possible in this study to determine possible adsorption sites as this requires simultaneous observation of atomically resolved adsorption atoms and the underlying clean surface at sub-monolayer coverages. However the observation of 4.9 \AA{} pentagonal clusters in the 1 ML film suggests that these features are a key building block of a quasicrystalline monolayer, whether on an icosahedral or a decagonal quasicrystalline substrate.

The saturation in coverage at 1 ML is similar to that found for Pb adsorption on $i$-Al-Pd-Mn \cite{Ledieu08}. There are a number of possible explanations for this unusual behaviour. There could be a vanishingly low binding energy for Pb on top of the quasiperiodic Pb monolayer, perhaps due to a one layer quantum size effect. Quantum size effects in growth have been previously observed for adsorption on quasicrystal surfaces \cite{Fournee05}. In a study of Ag growth on Fe(100) for films up to $N=15$ monolayers, films of $N=1,2$ and 5 monolayer thicknesses were found to have exceptional structural stability \cite{Luh01}. This possibility could be investigated using density functional theory calculations of the adsorption of Pb atoms on a Pb monolayer. Such calculations are now quite feasible \cite{Krajci05,Krajci06b,Krajci06c}.

A second explanation is that the effect could be due to some kinetic limitation  on the dissipation of the kinetic energy of the adsorbing Pb atoms leading to a vanishing sticking coefficient for Pb on the quasiperiodic Pb monolayer. A low sticking coefficient could result from a high phonon frequency for energy transfer from impinging Pb atoms to the quasicrystalline surface Pb layer \cite{Ledieu08}. Further work is needed before this question can be resolved.

There are also some differences between the $i$-Al-Pd-Mn/Pb and \mbox{$d$-Al-Ni-Co/Pb} systems.  During growth extended Pb islands are observed on $d$-Al-Ni-Co, whereas on $i$-Al-Pd-Mn/Pb a dispersed network of pentagonal clusters is clearly seen. This  indicates higher mobility for adsorbing atoms on the former substrate.  Although annealing to 673 K improves the order of a $i$-Al-Pd-Mn/Pb monolayer, annealing to 600 K of a $d$-Al-Ni-Co/Pb monolayer results in some desorption of Pb, leaving a porous, though well-ordered monolayer. Both of these observations suggest that the substrate/adsorbate interaction is reduced from that for $i$-Al-Pd-Mn/Pb  \cite{Ledieu08}.

\section{Conclusions}

Lead has been found to form a well-ordered quasiperiodic overlayer on the ten-fold surface of the decagonal Al-Ni-Co quasicrystal. This work extends the number of known systems where an adsorbate adsorbs pseudomorphically on a quasicrystal substrate. It is now apparent that chemistry is one of the most important factors in determining adsorbate structure: the elements which form pseudomorphic quasiperiodic overlayers are clustered in Groups III-V of the periodic table \cite{Smerdon08-PM}. Further structural studies are needed to determine the structure of these overlayers; they present a unique challenge to current quantitative surface structural techniques.

\section{Acknowledgments}

The UK Engineering and Physical Sciences Research Council (Grant number EP/D05253X/1) and the European Union Network of Excellence ``Complex metallic alloys'' Grant number  NMP3-CT-2005-500145 are thanked for financial support. Ian Fisher (Stanford University) is thanked for his help in growing the sample. We are grateful to Jim Evans (Ames Laboratory) for helpful discussions.


\newpage

\end{document}